\documentclass[%
reprint,
showpacs,
amsmath,amssymb,
aps,prl,
]{revtex4-1}
\usepackage{braket}
\usepackage{float}
\usepackage{graphicx}
\usepackage{dcolumn}
\usepackage{bm}
\usepackage{subfigure}
\usepackage{color}
\usepackage{ulem}

\begin{document}
\hyphenation{eq-ua-tions diff-er-ent only sce-nario also however equi-librium fila-ment results works re-mains two still account University} 

\title {Reply to ``Absence of Evidence for the Ultimate Regime in Two-Dimensional Rayleigh-B\'enard Convection"}

\author{Xiaojue Zhu$^{1,2}$}
\author{Varghese Mathai$^{1,3}$}

\author{Richard J. A. M. Stevens$^{1}$}

\author{Roberto Verzicco$^{4,1}$}

\author{Detlef Lohse$^{1,5}$}
\affiliation{$^1$Physics of Fluids Group and Max Planck Center for Complex Fluid Dynamics, MESA+ Institute and J. M. Burgers Centre for Fluid Dynamics, University of Twente, P.O. Box 217, 7500AE Enschede, The Netherlands\\
$^2$Center of Mathematical Sciences and Applications, and School of Engineering and Applied Sciences
Harvard University, Cambridge, Massachusetts 02138, USA\\
$^3$School of Engineering, Brown University, Providence, Rhode Island 02912, USA\\
$^4$Dipartimento di Ingegneria Industriale, University of Rome `Tor Vergata',
Via del Politecnico 1, Roma 00133, Italy\\
$^5$Max Planck Institute for Dynamics and Self-Organization, 37077 G\"ottingen, Germany}







\maketitle

In their Comment \cite {doe19} Doering {\it et al.} question our numerically found \cite{zhu2018b} onset of a transition to the ultimate regime 
of 2D Rayleigh-B\'enard convection. We disagree with their reasoning. 

\begin{figure}
    {\includegraphics[width=0.48\textwidth]{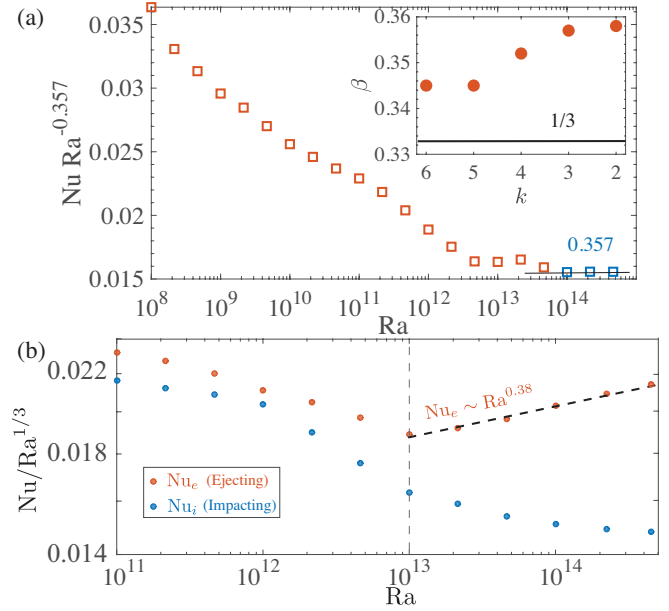}}
         \caption{
  (a) Nusselt number compensated by $\gamma = 0.357$, i.e.\
   a scaling exponent larger than the $\gamma = 1/3$ necessary  to prove the transition.  
  We took $\gamma = 0.357$ for the compensated plot as with that value the last three  data points show a 
   plateau. The error bars for the data are smaller than the symbols.
   The inset shows the effective scaling exponent $\gamma$, obtained from a 
   power law, fits $Nu \sim Ra^\gamma$  to the last $k$ data points in the main figure.
   It is always larger than 1/3, no matter how one interprets the data. 
    (b) The local heat transfer 
  in the plume emitting region (with effective slope 0.38) and in the plume impacting region. 
}
\label{fig1}
\end{figure}

To irrefutably settle the issue, we have extended our numerical simulations of ref.\ \cite{zhu2018b} to even larger $Ra$, namely now up to $Ra = 4.64 \times 10^{14}$,
sticking to the same strict numerical resolution criteria of both boundary layer (BL) and bulk.
The simulation at the highest $Ra$ was performed with a grid resolution of $31200\times25600$ with 28 points in the boundary layer. 
The evidence for the transition to the ultimate regime remains overwhelming:

\begin{enumerate}
\item
On the global heat transfer: 
$Nu(Ra)$, compensated with $Ra^{0.357}$,  is shown in figure \ref{fig1}a. 
An objective least squares fit of an effective power law $Nu \sim Ra^\gamma$ 
to the last 6 data points gives a  scaling exponent $\gamma = 0.345$; the last 5 data points give $\gamma = 0.345$, last four data points give $\gamma = 0.352$, last 3 data points give $\gamma = 0.357$, and last 2 points give $\gamma = 0.358$; i.e. no matter how the data is interpreted, the scaling exponent is {\it always} larger  than 1/3 and  monotonously increasing with $Ra$. 
\item 
The  key part of Ref.\ \cite{zhu2018b}  deals with {\it local} properties of the flow, see figures 2-4 of that paper. 
For the local heat flux  in the plume ejecting regime, beyond $10^{13}$ the effective scaling exponents is close to 
$\gamma = 0.38$, see figure \ref{fig1}. 
In contrast, it remains $\le 1/3$ in the plume impacting regime, which therefore with increasing $Ra$ 
looses more and
more relevance for the overall heat transfer. 
\item 
Beyond $10^{13}$, the horizontal velocity profiles $u^+ (y^+)$ in the BLs become logarithmic (see figure 2 of the Ref.\ \cite{zhu2018b}), signaling a turbulent BL, which is 
characteristic for the ultimate regime (as a presumption to derive the ultimate regime scaling in refs.\
 \cite{kraichnan1962,grossmann2011}), rather than one of laminar type as in the classical regime. 
\item 
 Finally, the transition to ultimate RB turbulence in the numerical data of 
\cite{zhu2018b} has also been confirmed through an extended self-similarity (ESS) analysis of
the temperature structure functions, see ref.\ \cite{krug2018}: 
In that paper we find no ESS scaling before the transition. However, beyond the transition and for large enough wall distance $y^+ >  100$, we find clear ESS behaviour, as expected for a scalar in a turbulent boundary layer.  
Therefore also that analysis  
provides strong evidence that the observed transition in the global Nusselt number around  $Ra =  10^{13}$ indeed is
the transition from a laminar type BL to a turbulent type BL. 
\end{enumerate}

\bibliography{RB_rough}

\begin{thebibliography}{5}%
\makeatletter
\providecommand \@ifxundefined [1]{%
 \@ifx{#1\undefined}
}%
\providecommand \@ifnum [1]{%
 \ifnum #1\expandafter \@firstoftwo
 \else \expandafter \@secondoftwo
 \fi
}%
\providecommand \@ifx [1]{%
 \ifx #1\expandafter \@firstoftwo
 \else \expandafter \@secondoftwo
 \fi
}%
\providecommand \natexlab [1]{#1}%
\providecommand \enquote  [1]{``#1''}%
\providecommand \bibnamefont  [1]{#1}%
\providecommand \bibfnamefont [1]{#1}%
\providecommand \citenamefont [1]{#1}%
\providecommand \href@noop [0]{\@secondoftwo}%
\providecommand \href [0]{\begingroup \@sanitize@url \@href}%
\providecommand \@href[1]{\@@startlink{#1}\@@href}%
\providecommand \@@href[1]{\endgroup#1\@@endlink}%
\providecommand \@sanitize@url [0]{\catcode `\\12\catcode `\$12\catcode
  `\&12\catcode `\#12\catcode `\^12\catcode `\_12\catcode `\%12\relax}%
\providecommand \@@startlink[1]{}%
\providecommand \@@endlink[0]{}%
\providecommand \url  [0]{\begingroup\@sanitize@url \@url }%
\providecommand \@url [1]{\endgroup\@href {#1}{\urlprefix }}%
\providecommand \urlprefix  [0]{URL }%
\providecommand \Eprint [0]{\href }%
\providecommand \doibase [0]{http://dx.doi.org/}%
\providecommand \selectlanguage [0]{\@gobble}%
\providecommand \bibinfo  [0]{\@secondoftwo}%
\providecommand \bibfield  [0]{\@secondoftwo}%
\providecommand \translation [1]{[#1]}%
\providecommand \BibitemOpen [0]{}%
\providecommand \bibitemStop [0]{}%
\providecommand \bibitemNoStop [0]{.\EOS\space}%
\providecommand \EOS [0]{\spacefactor3000\relax}%
\providecommand \BibitemShut  [1]{\csname bibitem#1\endcsname}%
\let\auto@bib@innerbib\@empty
\bibitem [{\citenamefont {Doering}\ \emph {et~al.}(2019)\citenamefont
  {Doering}, \citenamefont {Toppaladoddi},\ and\ \citenamefont
  {Wettlaufer}}]{doe19}%
  \BibitemOpen
  \bibfield  {author} {\bibinfo {author} {\bibfnamefont {C.~R.}\ \bibnamefont
  {Doering}}, \bibinfo {author} {\bibfnamefont {S.}~\bibnamefont
  {Toppaladoddi}}, \ and\ \bibinfo {author} {\bibfnamefont {J.~S.}\
  \bibnamefont {Wettlaufer}},\ }\href@noop {} {\bibfield  {journal} {\bibinfo
  {journal} {Phys. Rev. Lett.}\ }\textbf {\bibinfo {volume} {123}},\ \bibinfo
  {pages} {259401} (\bibinfo {year} {2019})}\BibitemShut {NoStop}%
\bibitem [{\citenamefont {Zhu}\ \emph {et~al.}(2018)\citenamefont {Zhu},
  \citenamefont {Mathai}, \citenamefont {Stevens}, \citenamefont {Verzicco},\
  and\ \citenamefont {Lohse}}]{zhu2018b}%
  \BibitemOpen
  \bibfield  {author} {\bibinfo {author} {\bibfnamefont {X.}~\bibnamefont
  {Zhu}}, \bibinfo {author} {\bibfnamefont {V.}~\bibnamefont {Mathai}},
  \bibinfo {author} {\bibfnamefont {R.~J. A.~M.}\ \bibnamefont {Stevens}},
  \bibinfo {author} {\bibfnamefont {R.}~\bibnamefont {Verzicco}}, \ and\
  \bibinfo {author} {\bibfnamefont {D.}~\bibnamefont {Lohse}},\ }\href@noop {}
  {\bibfield  {journal} {\bibinfo  {journal} {Phys. Rev. Lett.}\ }\textbf
  {\bibinfo {volume} {120}},\ \bibinfo {pages} {144503} (\bibinfo {year}
  {2018})}\BibitemShut {NoStop}%
\bibitem [{\citenamefont {Kraichnan}(1962)}]{kraichnan1962}%
  \BibitemOpen
  \bibfield  {author} {\bibinfo {author} {\bibfnamefont {R.~H.}\ \bibnamefont
  {Kraichnan}},\ }\href@noop {} {\bibfield  {journal} {\bibinfo  {journal}
  {Phys. Fluids}\ }\textbf {\bibinfo {volume} {5}},\ \bibinfo {pages} {1374}
  (\bibinfo {year} {1962})}\BibitemShut {NoStop}%
\bibitem [{\citenamefont {Grossmann}\ and\ \citenamefont
  {Lohse}(2011)}]{grossmann2011}%
  \BibitemOpen
  \bibfield  {author} {\bibinfo {author} {\bibfnamefont {S.}~\bibnamefont
  {Grossmann}}\ and\ \bibinfo {author} {\bibfnamefont {D.}~\bibnamefont
  {Lohse}},\ }\href@noop {} {\bibfield  {journal} {\bibinfo  {journal} {Phys.
  Fluids}\ }\textbf {\bibinfo {volume} {23}},\ \bibinfo {pages} {045108}
  (\bibinfo {year} {2011})}\BibitemShut {NoStop}%
\bibitem [{\citenamefont {Krug}\ \emph {et~al.}(2018)\citenamefont {Krug},
  \citenamefont {Zhu}, \citenamefont {Chung}, \citenamefont {Marusic},
  \citenamefont {Verzicco},\ and\ \citenamefont {Lohse}}]{krug2018}%
  \BibitemOpen
  \bibfield  {author} {\bibinfo {author} {\bibfnamefont {D.}~\bibnamefont
  {Krug}}, \bibinfo {author} {\bibfnamefont {X.}~\bibnamefont {Zhu}}, \bibinfo
  {author} {\bibfnamefont {D.}~\bibnamefont {Chung}}, \bibinfo {author}
  {\bibfnamefont {I.}~\bibnamefont {Marusic}}, \bibinfo {author} {\bibfnamefont
  {R.}~\bibnamefont {Verzicco}}, \ and\ \bibinfo {author} {\bibfnamefont
  {D.}~\bibnamefont {Lohse}},\ }\href@noop {} {\bibfield  {journal} {\bibinfo
  {journal} {J. Fluid Mech.}\ }\textbf {\bibinfo {volume} {851}},\ \bibinfo
  {pages} {R3} (\bibinfo {year} {2018})}\BibitemShut {NoStop}%
\end{thebibliography}%

\end{document}